# Unipolar resistive switching in cobalt titanate thin films

ATUL THAKRE[1,2], A. K. SHUKLA[1], R. S. KATIYAR[3] and ASHOK KUMAR[1,2]

[1] *CSIR-National Physical Laboratory, Dr. K. S. Krishnan Marg, New Delhi 110012, India*
[2] *Academy of Scientific and Innovative Research (AcSIR), CSIR-National Physical Laboratory (CSIR-NPL) Campus, Dr. K. S. Krishnan Road, New Delhi 110012, India*
[3] *Department of Physics, and Institute for Functional Nanomaterials, University of Puerto Rico, San Juan, PR 00931-3343, USA*

PACS 73.40. Rw – Metal-insulator-metal structures

PACS 77.80. Fm – Switching phenomena

PACS 77.55.-g – Dielectric thin films

**Abstract –** We report giant resistive switching of an order of $10^4$, long-time charge retention characteristics up to $10^4$ s, non-overlapping SET and RESET voltages, ohmic in low resistance state (LRS) and space charge limited current (SCLC) mechanism in high resistance state (HRS) properties in polycrystalline perovskite Cobalt Titanate ($CoTiO_3$ ~ CTO) thin films. Impedance spectroscopy study was carried out for both LRS and HRS states which illustrates that only bulk resistance changes after resistance switching, however, there is a small change (<10% which is in pF range) in the bulk capacitance value in both states. These results suggest that in LRS state current filaments break the capacitor in many small capacitors in a parallel configuration which in turn provides the same capacitance in both states even there was 90 degree changes in phase-angle and an order of change in the tangent loss.

**Introduction.** In the recent years, Resistive Switching (RS) behaviour in binary and ternary oxide thin films has been studied for their application as Non-volatile Random Access Memory (NV-RAM). In the RS phenomena, the resistance of the device drastically changes due to change in external applied electric field with suitably applied compliance current. It promises various advantages over the existing NVRAM technologies such as extremely low power consumption, greater endurance, high charge retention capability and the most important is ultimate atomic scalability[1-6]. There are mainly two types of resistive switching phenomenon occur in the system depending on an application of an external electric stimulus; when SET and RESET processes occurs in the same polarity is called unipolar, whereas if SET and RESET process occurs on the opposite electric (E)-field polarity is termed as bipolar resistive switching. A comparatively large high resistance state (HRS) to low resistance state (LRS) $R_{HRS}/R_{LRS}$ ratio makes the unipolar resistive switching preferable over the bipolar resistive switching because of easily WRITE and READ process of the data. A simple crossbar structure of RS devices has proved itself a promising candidate for the next generation NVRAM elements[7, 8]. Most of the organic and inorganic compounds, especially binary transition metal oxides such as $TiO_2$[9], $NiO$[10, 11], $WO_x$, $Cu_2O$[12], $ZrO_2$[13] and some other binary oxides such as $HfO_x$[14, 15], $Al_2O_3/WO_x$[16] and etc. have shown resistive switching phenomena, however studies on these systems are in nascent phase. In general, mostly bipolar RS behaviour has been reported in the various complex perovskite type of compounds thin films such as Fe doped $SrTiO_3$[17], Cr doped $SrZrO_3$[18], $(Ba,Sr)TiO_3$[19] etc. There are few transition metal oxides and perovskite types of compounds also exist in nature which show unipolar resistive switching such as ZnO[20], $SnO_2$[21], NiO[22] $BiFeO_3$[23], $SrTiO_3$[24].

In this letter, we report reproducible unipolar resistive switching behaviour in the polycrystalline Cobalt Titanate ($CoTiO_3$ ~ CTO) thin films deposited on $Pt/TiO_2/SiO_2/Si$ substrate by pulsed laser deposition technique. We employed impedance spectroscopy technique to investigate the formation of current filaments during LRS state.

**Experimental.** – The $CoTiO_3$ target was prepared by the conventional solid-state reaction method using high purity $Co_3O_4$ (99.9%) and $TiO_2$ (99.9%) precursors. These precursors were thoroughly mixed and calcined at 1050 $^0$C for 5 hours, after confirmation of phase formation; one-inch circular target was prepared at 1250 $^0$C for 8 hours of heat treatment. The CTO films were deposited on $Pt/TiO_2/SiO_2/Si$ substrate by pulsed laser deposition (PLD) technique at growth temperature 750 °C, oxygen pressure of 5 mTorr and laser energy density 1.5 $J/cm^2$. The average thickness of the films was measured by profilometer and found in the range of 300 nm. X-ray diffraction with monochromatic source Cu-$K_\alpha$ radiation (BRUKER D8 ADVANCE) and atomic force microscopy studies were performed on the samples to check the phase purity, and crystalline quality, respectively. The top Au electrode was deposited using shadow mask of diameter 200 µm and thickness ~ 80 nm using dc sputtering. The Agilent (model B2901a) & Keithley 236 source meter with micro probe station were employed to carry out current-voltage (I-V) and retention measurement. An Impedance Analyser Hioki 3532-50 LCR HiTESTER with an oscillating voltage of 0.5 V and frequency 100 Hz to 1 MHz was used to measure the resistance and capacitance for LRS and HRS states.

The phase purity and crystalline quality of the CTO thin films on $Pt/TiO_2/SiO_2/Si$ substrate were studied by x-ray diffraction (XRD) patterns. Fig. 1 shows broad peaks at 32.8 and 35.4 Bragg's angle which belongs to rhombohedral structure matched with JCPDS file no 77-1373 and bulk CTO target (left-hand side inset Fig.1). The substrate peaks are marked in Fig.1 which dominates XRD pattern of CTO films [25]. The Bragg's angle (2ϴ) at 32.8 and 35.4 belong to CTO thin film and peak at 33.4 Braggs angle belongs to platinised silicon substrate. All three peaks together provide a broaden peak which is hard to distinguish. The CTO thin film illustrates broaden XRD peaks compared to bulk that favor the presence of nanoscale grains size. The average grains sizes are in the range of 20-40 nm as can be seen in AFM images (right inset Fig.1) with average surface roughness nearly 5-7 nm.

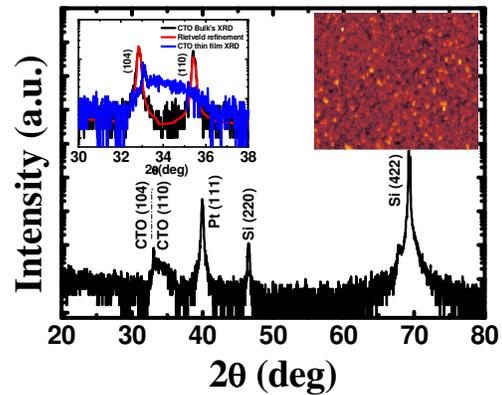

Fig.1: XRD patterns of the CTO thin film on $Pt/TiO_2/SiO_2/Si$ substrate. In inset, on the left side XRD peaks of CTO target with rietveld refinement fitting and film were compared; whereas on the right side, surface topography image of the film surface is shown.

**Results and Discussion.** – As grown CTO thin films show very high resistance in the range of Giga-ohms. To find the first SET voltage, a positive bias is applied to the top electrode with compliance current ($I_{cc}$ ~ 2 mA). Most of the devices show first SET voltages in the range of 8-10 V, where current abruptly changes from several nA to 2 mA (Current compliance in this case). This process is called filament formation and the device got switched from HRS to LRS (SET process). After the occurrence of SET process, the current compliance value is increased to a higher value (~100 mA) to make the filament ruptured and hence RESET the device from LRS to HRS state. Fig. 2(a) shows that when a small positive bias

(~3V) was applied; the current value drops abruptly which means the device resistance goes from LRS to HRS (RESET process).

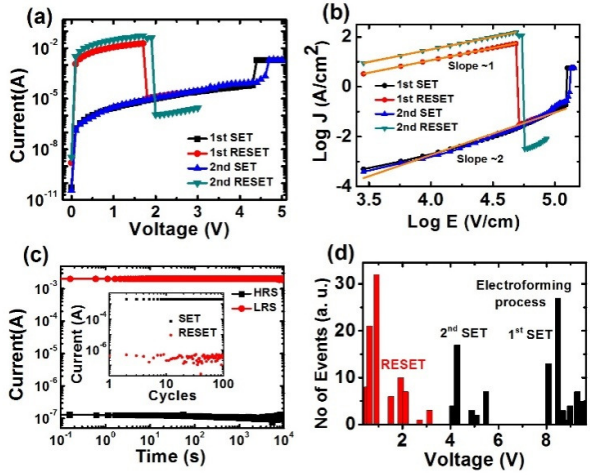

Fig. 2: (a) Current-voltage graph for the unipolar resistive switching, (b) Ohmic and SCLC conduction mechanisms fitting for the I-V data, (c) The devices retention characteristics in the HRS and LRS over $10^4$ seconds with a bias voltage of 1 V and compliance current of 2 mA (the endurance characteristics of the device for 100 cycles have been shown in the inset), (d) A histogram distribution of SET and RESET voltages have been shown.

Thus, the high current compliance limits the passing of very high current through the conducting filament due to local heating near the electrode-electrolyte interface. Here in this situation, it causes the rupture of the conducting filament formed in the electroforming process (SET process) due to the Joule Heating [26]. In the Fig 2(d), it can be seen that the 1st SET process is occurring near ~8V, while the next SET occurs at comparatively lower voltage ~4.5V. It indicates that the conducting filament was not completely ruptured in the 1st RESET process and there may be some residual conducting path present in the system, which causes a lower SET voltage for the device as compared to the pristine CTO thin film. The CTO devices show a SET and RESET voltage window depending upon the device surface topography and the residual left in the conducting path. To RESET the device, the $I_{CC}$ was fixed at around 100 mA and then applied voltage was increased gradually. At ~2 V, the device RESETs to high resistance state. This SET and RESET process was repeated multiple times and we found that the SET voltage was varying in between 4-5.5 V and RESET phenomena was rather occurring below 2 V. These SET and RESET voltage values also have a good non-overlapping window of around 3 V. The resistance ratios of HRS to LRS were measured around $10^4$ at lower applied electric field (~ 0.1 V), which offers an ease of the reading of both the resistive states. A histogram of SET and RESET voltages has been shown in the Fig 2(d). In every pristine CTO resistive device, the very first SET process takes place in the range of 8.5 to 10 V and then after the first RESET operation, SET process occurred at comparatively lower voltage in the range of 4 to 5.5 V. Whereas the RESET operation occurred mostly in the range of 0.5 to 2.5 V. We observed these many multiple SET and RESET operations as shown in histogram. It was also found that the SET and RESET voltages also depend on the time interval (few seconds to milliseconds) between two consecutive applied voltages i.e. with the same compliance current value, SET and RESET will occur at lower voltage with longer time interval of applied voltage. These data are also included in the histogram.

For application point of view, retention and endurance characteristics are two main degrees of merits for resistive switching memories. To check these properties, retention test of the device in both HRS and LRS was performed under a constant voltage of 2V as shown in Fig 2(c). These devices show negligible deterioration in charge retention capacity over a period of more than 10000 seconds confirming ability to hold the charge for long time in both the states (HRS and LRS). While for the endurance characteristics test shown in the inset of the Fig 2(c), the device was examined for 100 cycles of continuous repeated SET and RESET processes and it shows almost constant SET and RESET current values. In conclusions, the device showed good repeatability and stability in the switching operations which is a suitable quality for non-volatile memory application.

The space charge limited conduction (SCLC) model was explored to understand the conduction mechanisms in the unipolar resistive switching system. The current-voltage graphs were re-plotted as log J vs. log E and obtained the magnitude of slopes in various E-field regions (Fig. 2 (b)). The measured slopes in the LRS were found nearly one which clearly confirms the ohmic conduction in the system due to the formation of current carrying path. However, in HRS the magnitude of slope is near to ~2 before the SET process. In the lower applied E-field region, i.e. before the switching and fast forward of charge carriers, devices show high resistance. In this applied E-Field region, the slope is ~2, which indicates trap-assisted space charge limited conduction (SCLC) [27]. The SCLC conduction phenomena can be classified into three categories depending on the magnitude of slopes obtained by linear fitting[28,29]. Initially, when the applied electric field is too small, there is a small number of energetic charge carriers present in the system causes a small conduction current, and hence device shows a high resistance. With increase in applied electric fields, some of the charge carriers injected across electrode-electrolyte interface and some of them developed in the films, among these charge carriers, some of them later captured by the traps and remaining contribute to the conduction process[30, 31]. This process is referred as trap limited SCLC. When the applied electric field is further increased, a large amount of charge carriers injected in the matrix and fill most of the trap centres. In this condition the injected charge carrier increases very rapidly and quickly filled the trap centres and developed a condition of trap-free conduction with a very narrow window of applied field. In this case, the conduction is fully dominated by these injected charge carriers. A relation between current density and the applied electric field is as follows;

$$J_{SCLC}= (9/8) \mu\varepsilon\theta \, E^2/d$$

where, $\mu$ is the mobility of charge carriers, $\varepsilon$ is static dielectric constant, $\theta$ denotes the ratio of injected free charge carriers to total charge carriers, E denotes applied electric field and d is thickness of the films. We also examined the other conduction mechanisms such as Schottky emission and Poole Frankel emission with linear fitting the ln J vs. $E^{1/2}$ graph and ln J/E vs. $E^{1/2}$ graph, respectively [32-34]. The magnitude of slopes obtained from linear fitting provides unrealistic optical dielectric constants, hence ruled out of its role in the conduction mechanism. Chae *et* al. [35] has reported the unipolar resistive switching in $TiO_2$ due to current-filament formation and it RESET due to Joule heating based giant enhancement in local temperature which can reach up to 800 K. Similar effects may prevailed in CTO where the unipolar resistive switching phenomena is due to electroforming of the conducting filaments and rupture due to Joule heating effect.

The impedance spectroscopy measurements were carried out in each resistive state to understand the microstructure and electrical property relation. Initially, when the device was in the HRS, the impedance spectroscopy [36] parameters such as impedance, dielectric loss, phase angle, resistance and capacitance with respect to the frequency were recorded. Fig 3(a) and (c) show the real and imaginary impedance variation as a function of frequency in the HRS and LRS, respectively. A drastic decrease in the magnitude of the real part of impedance and a giant shift in relaxation frequency (peak $f_{max}$) towards higher frequency side were observed with a resistance switching from HRS to LRS.

Fig 3(b) and (d) illustrate the Nyquist plot (Z´´ vs Z´) in HRS and LRS, respectively; the experimental data were fitted with the equivalent electrical circuit comprising a parallel combination of a bulk resistance ($R_{bulk}$) and bulk capacitance ($C_{bulk}$) with the series connection of a contact resistance ($R_s$). This experimental data fitted well with the equivalent electrical circuit model (solid line fitting) that suggest in any condition the bulk resistance of the system changes at least three order of magnitude from HRS to LRS, however the overall bulk capacitance remains almost same in both states. The various fitted parameters values of $R_s$, $R_{bulk}$, and $C_{bulk}$ for HRS are 104.2 $\Omega$, 6.8446

MΩ, & 2.09E-10 F whereas for LRS are 33.38 Ω, 49777 Ω, & 1.903E-10 F, respectively.

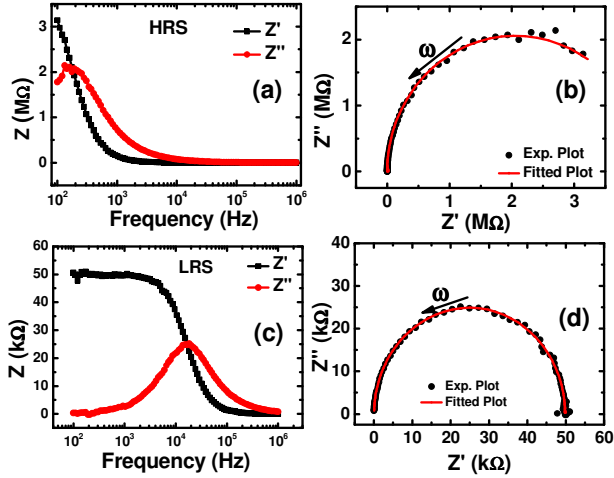

Fig. 3: (a) and (c) Real and imaginary part of impedance spectra with respect to frequency for HRS and LRS, respectively, (b) & (d) Nyquist plot and equivalent circuit model fit for the HRS and LRS, respectively.

These results gave us a clear vision about the resistive switching. During RS process, only resistance changes, however, there is a very small change (~ <20 pF or less than 10%) in capacitance. It means that during HRS to LRS switching process, several current carrying path (current filaments in this case) forms in the matrix and the capacitor (metal-insulator-metal) physical structure divided into several capacitors connect in parallel so the resultant capacitance remains almost same even the phase change by 90 degree and tangent loss by x10 times. The relaxation time ($\omega_{max}\tau=1$) of charge carriers also changes five to six orders of magnitude after switching, i.e. it changes from seconds to several micro-seconds (~ μs) from HRS to LRS. These results suggest that charge carriers very quickly hop from one state to another state or move from one position to another position and significantly increase the conduction of the system in LRS.

The capacitance and tangent loss of devices in both HRS and LRS states over the frequency range of 100 Hz to 1 MHz are shown in Fig. 4(a) and (b), respectively. The capacitance of the pristine device is almost having the same magnitude within the experimental limitations and errors. The bulk capacitance is almost same in each RS states with a tangent loss at least ten times high in LRS. The magnitude and behavior of dielectric constant and loss suggest that capacitor structure remains intact with the formation of current filaments in the matrix.

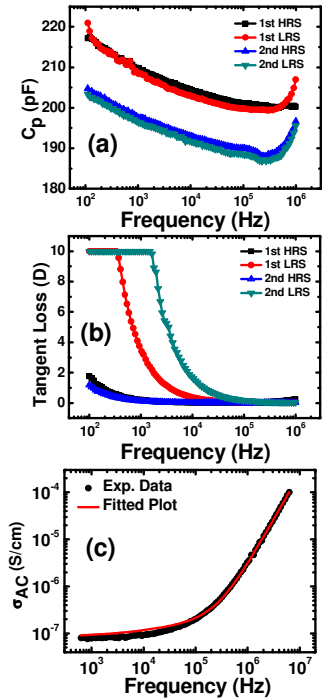

Fig. 4: (a) Frequency dependant plot of capacitance in both HRS and LRS, (b) Frequency dependant plot of tangent loss and (c) ac conductivity as a function of frequency for HRS and their power law fitting.

The ac conductivity ($\sigma_{AC}$) of CTO thin films in HRS (LRS is ohmic) were obtained using the empirical relation given below,

$$\sigma_{AC} = \omega \varepsilon_0 \varepsilon_r \tan\delta$$

where, $\omega$ is angular frequency, $\varepsilon_0$ is absolute permittivity, $\varepsilon_r$ is relative permittivity and $\tan\delta$ is the dissipation factor. The ac conductivity as a function of frequency is shown in the Fig.4(c). In general, a simple Jonscher power law relation is the ideal model to define the frequency dependent behaviour of ac conductivity. Whereas in this system, there are several mechanisms which are contributing in the conductivity, so the simple Jonscher power law does not fit well with our experimental data. A modified Jonscher power double law fits well to explain our experimental data;

$$\sigma_{AC} = \sigma_0 + A\omega^p + B\omega^q$$

where, $\sigma_0$ is frequency independent conductivity, $\omega$ is the angular frequency and A & B are constants, and p, q are exponents [37, 38]. In low-frequency regions (<10 kHz), almost frequency independent long-range charge carriers contribute to conductivity. The intermediate frequency region represents competing translational hopping of charge carriers with exponent term p lies between 0 and 1. However, the high-frequency regions (>100 kHz) relates to localized or re-orientationally hopping of charge carriers with a magnitude of exponent lies between 1 and 2. It happens due to very large relaxation time (~ seconds) in HRS and occurrence of unsuccessful hoping among the nearest neighbour.

**Conclusion. –** In conclusion, a robust and reproducible unipolar resistive switching was obtained in the polycrystalline CTO thin films. The crossbar metal-CTO-metal structure illustrates distinct resistive states with $10^4$ $R_{OFF}/R_{ON}$ resistance ratio and excellent capacity to hold the charge for a long time suggests strong possibility as NVRAM memory elements applications. The devices have shown excellent switching properties with small SET and RESET voltage window favouring design of distinguishable logic and bit states. The trapped charge assisted SCLC conduction mechanism guide the SET and RSET process. An impedance spectroscopy analysis provides the similar magnitude of capacitance in both LRS and HRS states with 90-degree change in phase angle and ten times increase in tangent loss suggesting the formation of current filaments with negligible effect on microstructure.

\*\*\*

A.K. acknowledges the CSIR-MIST (PSC-0111) and CSIR-Mission project for the financial assistance. Atul Thakre would like to acknowledge the CSIR (JRF) to provide fellowship to carry out Ph. D program. Authors would like to thank Mr. Ravikant for making PLD target and pellets, Dr. Ritu Srivastav for surface topography, Dr. Ranjana Mehrotra (Head, P.M.M.), & Dr. Sanjay Yadav, for their constant encouragement and support.